# Understanding Emojis 🙂 in Useful Code Review Comments

Sharif Ahmed
sharifahmed@u.boisestate.edu
Boise State University
Boise, Idaho, United States

Nasir U. Eisty
nasireisty@boisestate.edu
Boise State University
Boise, Idaho, United States

## ABSTRACT

Emojis and emoticons serve as non-verbal cues and are increasingly prevalent across various platforms, including Modern Code Review. These cues often carry emotive or instructive weight for developers. Our study dives into the utility of Code Review comments (CR comments) by scrutinizing the sentiments and semantics conveyed by emojis within these comments. To assess the usefulness of CR comments, we augment traditional 'textual' features and pre-trained embeddings with 'emoji-specific' features and pre-trained embeddings. To fortify our inquiry, we expand an existing dataset with emoji annotations, guided by existing research on GitHub emoji usage, and re-evaluate the CR comments accordingly. Our models, which incorporate textual and emoji-based sentiment features and semantic understandings of emojis, substantially outperform baseline metrics. The often-overlooked emoji elements in CR comments emerge as key indicators of usefulness, suggesting that these symbols carry significant weight.

## 1 INTRODUCTION

The Modern Code Review (MCR) process is embraced by both industry and open-source developers, with organizations customizing it to suit their specific needs. This paper focuses on the last phase of MCR, review feedback, more specifically, CR comment, which is empirically found important in both open-source [4, 10, 19] and industry [4, 5, 14]. Unfortunately, a study [5] at Microsoft revealed that 34.5% of the CR comments are not *useful*. To tackle the not-useful CR comments, researchers have defined the usefulness of CR comments, mined and annotated datasets of **useful** CR comments [15, 17, 19], studied developers' perceptions in commercial [5] and open-source projects [19], analyzed factors from various aspects, and used machine learning classifiers to automatically predict the usefulness of CR comments [1]. However, CR comments are written in verbal (i.e., natural language text) and/ or non-verbal (i.e., emoji, emoticons, animation, or votes) forms.

Lu et al. [13] reported a notable uptick in the use of emojis in pull request comments, from less than 1% to 10%, over half a decade. In a separate empirical study, Park and Sharif [16] found that developers pay significantly more attention to emojis compared to the body text in CR comments. Furthermore, a study by Wang et al. [20] suggested that emoji reactions can positively influence collaboration in the code review process. Recently, Lu et al. [12]

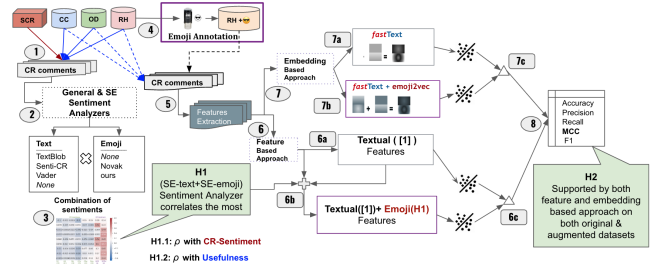

**Figure 1: Overview of Our Approach to Hypotheses Testing**

employed emojis as predictors to forecast the dropouts of remote developers on GitHub successfully. Interestingly, they found that developers who did not use emojis were three times more likely to drop out in the subsequent year than those who did.

Conversely, Ahmed and Eisty [1] delved into the effectiveness and comprehension of CR comments by examining factors like sentiment, polarity, and formality within the text. However, existing research has not yet explored the role of non-verbal elements such as emojis or emoticons in the context of useful CR comments. Given that these non-verbal cues can convey sentiments or polarity much like their verbal counterparts, our initial hypothesis is **H1: "The sentiments expressed through emojis, in conjunction with those in CR comment texts, are correlated with the overall sentiment (H1.1) or usefulness (H1.2) of Code Reviews"**.

These non-verbal cues do more than convey sentiment or add visual flair; they also serve as instructional elements for developers [18]. Therefore, these cues could influence the perceived usefulness of CR comments, either emotionally or technically. Building on this, our next hypothesis is **H2: "Incorporating emojis alongside text and semantic content will improve the prediction of the usefulness of CR comments"**.

We aim to examine these hypotheses by utilizing sentiment analyzers and pre-trained models for both text and emojis (see Figure 1). Given the absence of specialized emoji sentiment analyzers for CR comments, we plan to annotate an existing dataset with emoji-related sentiments. Additionally, we intend to enrich the same dataset comprehensively with emojis to achieve a more holistic understanding. Our findings indicate that general emoji sentiments offer valuable insights, while domain-specific emoji sentiments provide even deeper context. Above all, the semantics of emojis prove to be the most informative factor in determining the usefulness of CR comments when utilizing existing datasets.

The primary contributions of this paper are: (1) a compilation of emoji-sentiment scores specific to CR comments, (2) a CR comment dataset annotated with emojis, and (3) empirical support for the significance of emojis in CR comments, derived from an analysis of their sentiment and features obtained from pre-trained embeddings.

## 2 METHODOLOGY

### 2.1 Emojis and their Features

Here, we describe how we extract emojis from CR comments and how we obtain features for emojis from existing tools and models.

*2.1.1 Emoji Extraction.* We convert various non-Unicode forms of emojis, including codepoints '*+U1F60A*', emoticons ':)', and emoji expressions ':smile:', into Unicode emoji characters '😊'. Specifically, we use *GroupMe Shortcuts*[1] to convert emoticons into emoji expressions. Following this, we utilize the *python package*[2] to transform all emoji expressions into their corresponding Unicode emojis.

*2.1.2 Features from Emoji.* Given that emojis convey emotions and various other cues, our initial focus is obtaining sentiment scores and their semantic features, as described below.

**i) Emoji Sentiment:** As emojis have expressions and descriptions in natural language, and considering that sentiment analyzers exist for natural language, we decode the emojis into their textual expressions. We then use TextBlob[3], a sentiment analyzer for natural language (i.e., English) text, to generate sentiment scores for these emoji expressions. However, this sentiment analyzer can not produce meaningful scores for the emojis from their expressions.

Next, we employ SEntiMoji [6], a sentiment analyzer fine-tuned on the DeepMoji [8], which incorporates relevant GitHub communication data, although it excludes CR comments. However, this tool is confined to the 64 emojis available in DeepMoji. But it only supports 5 out of the 14 Unicode emojis in the RevHelper [17] dataset. Given this limitation, we use the emoji sentiment scores provided by Novak et al. [11], which covers around 80% of the emojis we are interested in. We refer to these scores as *general emoji sentiment*. Since none of these sentiment sources are tailored for CR comments, we plan to manually annotate the sentiment scores specifically for emojis found in CR comments. This annotation process is outlined in Section 2.2.1, and we label these as *code-review (domain-specific) emoji sentiment*. In cases where a single CR comment contains multiple emojis, we aggregate the sentiment scores to derive the overall sentiment value for the emojis in that CR comment.

**ii) Emoji Embedding:** In our quest to identify useful CR comments, we opt for a semantic approach using transfer learning embedding techniques, which offer a wealth of information. Specifically, we employ an emoji pre-trained model, *emoji2vec* [7], for this purpose. This model has been trained on 6088 descriptions from the Unicode emoji standard, covering 1661 emoji symbols, and is built upon the Google News word2vec model. Using this pre-trained model, we then aggregate these emoji embeddings for each CR comment. We notice Unicode 🙍 is missing in both emoji sentiment [11] and pre-trained emoji embedding [7] besides GitHub's customized 🐿 and 👹.

### 2.2 Emoji Augmentation

Given that the datasets are somewhat outdated, the limited presence of emojis might not sufficiently aid the classification process. Initially, we turn to SEntiMoji [6] to populate the CR comments with emojis. However, Table 1 shows it overwhelmingly recommend the 💯 emoji for 86-94% of the CR comments in the studies [15, 17, 19]. Additionally, we observe that other suggested emojis are often not contextually appropriate. For instance, SEntiMoji recommends the 💙 emoji for CR comments like '_exit (EXIT_FAILURE),' which seems mismatched. Given these challenges, we identify two options: mine new CR comments from repositories to create and annotate a fresh dataset for usefulness or add emoji annotations to an existing dataset. Due to the scope of this study, we opt for the latter approach. We select the RevHelper [17] dataset for this purpose, as it comes with fully manually annotated CR comment usefulness. With this chosen dataset [17], we contribute (i) emoji sentiments and (ii) augment the RevHelper [17] by adding emojis to CR comments. The following text elaborates on this process.

*2.2.1 Emoji Sentiment Annotation.* Given that none of the existing emoji sentiment scores are tailored to the code review domain, we experiment with manually annotated sentiment scores[4]. This approach is inspired by a relevant blog article[5].

*2.2.2 Emoji Annotation.* Since the selected dataset [17] does not include corresponding code changes, our first step is to establish guidelines based on existing empirical studies that have mapped the intentions and usages of emojis in different contexts such as issues, pull requests, and comments [13, 18]. These studies provide insights into emoji usage across various GitHub activities, including READMEs. Additionally, we consult two popular GitHub-based guidelines to enrich our understanding: **gitmoji**[6], which has garnered 14,900 stars and 803 forks and provides a comprehensive list of emojis for use in commit messages, and **CREG**[7], with 412 stars and 26 forks, which offers a guideline for emoji use in code reviews. After synthesizing information from these resources, we compile a list of emojis and their likely contexts within CR comments, shared in our artifact[4]. Note that we opt for the Unicode emojis 🚢 🚀 and 😉 as substitutes for custom GitHub emojis like '🐿:shipit: (squirrel)' and '👹:trollface: (ogre, mischievous smile)' since the latter is not directly available in Unicode format. Finally, guided by these compiled resources, we simulate the context that a code reviewer might find themselves in and proceed to annotate the emojis on RevHelper [17] train-test dataset accordingly. We name this as *RH+*😊 across the paper.

The authors disagree on several annotations, such as `Do you want a space after the comma?`. Here, one author label 🤔 assumes that the reviewer shared a thought, while another annotator label with 🔨 assumes that the reviewer suggested an update. Though the annotation is very subjective, we find an inter-annotator agreement score of 0.81 (**Cohen's Kappa**), which is an almost perfect agreement. Next, we resolve the remaining disagreements

---

[1]https://support.microsoft.com/en-us/office/emoticon-keyboard-shortcuts-5dbe678c-cef7-4a63-aa62-f07c2f38b267
[2]https://pypi.org/project/emoji/
[3]https://pypi.org/project/textblob/
[4]**Our Artifact:** https://doi.org/10.5281/zenodo.10552771
[5]https://blog.carbonfive.com/how-to-use-actionable-emojis-in-your-pull-request-reviews
[6]https://github.com/carloscuesta/gitmoji
[7]https://github.com/erikthedeveloper/code-review-emoji-guide

Table 1: Emojis in Useful CR comments

| Dataset | Size | Emojis in the Dataset | SEntiMoji's [6] Emoji Prediction |
|---|---|---|---|
| RevHelper [17] | 1,481 | 👍(18), 😅(9), 😒(3), 😊(2), 😳(1), 👎(1), 😮(1), 🙍(1), 😛(1), 🙎(1), 🕵(1), 💩(1) | 💯(1268), 😣(180), 😦(22), 😫(9), ❤(2) |
| Chromium Conv. [15] | 3,794 | 😄(54), 😅(6), 😒(3), 🤓(2), 😕(1), 😊(1), 👣(1) | 💯(3572), 😣(166), 😦(42), 😫(9), 💙(2), ❤(2), 💩(1) |
| OpenDev [19] | 2,654 | 😄(38), 😒(8), 😛(3), 👍(2), 😉(3) | 💯(2232), 😣(220), 😦(23), ❤(11), 😫(8), 🙈(2), 💙(1), 😿(1), 😒(1), ✨(1) |

Here, numbers inside the parentheses following the emojis denote the appearance of the emojis through face-to-face discussion. Lastly, we analyze and observe that the top 10 emojis are: ⛏(527), 🤔(323), 👀(129), 👎(107), 👍(97), 😦(86), 💡(73), 😄(63), 🚨(62), and 😉(32).

## 2.3 Experiment Setup

*2.3.1 Verbal and/or Non-verbal Sentiments.* To validate our first hypothesis, **H1**, we conduct sentiment analysis on CR comments and their contained emojis separately. We then examine the Pearson correlation ($\rho$) between these sentiments with and without emojis and between code review sentiment and the perceived usefulness of CR comments. We use previously annotated datasets [2, 15, 17, 19] to gauge the sentiment and usefulness of code reviews. Specifically, we employ TextBlob's polarity score ($G_T$) and Senti-CR ($CR_T$) [2] to analyze the sentiments expressed in CR comments, respectively, from general and code-review-specific perspectives. For the sentiment analysis of emojis, we use Novak's emoji sentiment scores ($G_E$) from the general domain [11] and our manually annotated emoji sentiment scores ($CR_E$) as detailed in Sec 2.2.1, which are specific to the context of code reviews. Additionally, we utilize VADER [9], ($G_{TE}$), another general-domain sentiment analyzer capable of interpreting emojis. Subsequently, we experiment with combinations of verbal and non-verbal sentiments extracted from general and software-specific domains to see how these elements interact and contribute to our understanding.

*2.3.2 Usefulness Prediction Considering Emojis.* To evaluate our second hypothesis, **H2**, we first acquire features related to emojis, including their sentiment scores and embeddings. We then explore the impact of these features on models designed to predict the usefulness of CR comments. Specifically, we experiment by including or excluding these emoji-related features from existing non-emoji-based prediction models.

**Feature-based.** We utilize all existing textual features from prior research [1]. We leverage the author-provided publicly available source codes to implement these features. The model comprising these features serves as the baseline for our feature-based approach. As the existing datasets lack developer experience and review activity records [1], we excluded the features from these non-textual aspects of features. Using these feature values, we then employ Random Forest classifiers to predict the usefulness of CR comments. We choose prior evaluation metrics [1] and Matthews Correlation Coefficient (MCC) to evaluate the models. To establish a baseline performance, we execute a stratified 10-fold cross-validation with existing textual features. Subsequently, we incorporate emoji sentiments selected based on the best combination identified in Section 2.3.1 into the same training and testing setup to obtain the performance of the proposed emoji-aware model.

**Embedding-based.** To our knowledge, no existing research has employed pre-trained models specifically for predicting the usefulness of CR comments. As a baseline for non-emoji embedding-based models, we utilize fastText [3]. For the emoji embedding-based model, we supplement fastText embeddings with those from emoji2vec [7]. In this case, we adhere to the same training and evaluation procedures as used in the feature-based approach, substituting hand-crafted features with embeddings.

**With Augmented Emojis.** Given the sparse presence of emojis in existing datasets (~2-3%), we gather all the CR comments featuring emojis to create a specialized dataset, which we call $D_{all}^{\in 🙂}$. Finally, we train and evaluate the aforementioned prediction models on this specialized dataset and our augmented dataset, $RH^{+🙂}$.

## 3 RESULTS AND DISCUSSION

### 3.1 Verbal and/or Non-verbal Sentiments (H1)

In our analysis, we explore the role of emojis in CR comments from two perspectives: sentiment and usefulness. Here, we exclude our emoji-augmented dataset, $RH^{+🙂}$.

In Figure 2, the rightmost columns indicate that general text ($G_T$) and emoji ($G_E$) sentiments do not significantly correlate with sentiments in CR comments where emojis are present in roughly 2-3% of the cases. However, CR comments that do contain emojis moderately correlate with domain-specific emoji sentiments ($CR_E$), weakly with general emoji sentiments ($G_E$), and have a negligible correlation with general text sentiment ($G_T$) scores. Intriguingly, the emoji-aware general sentiment ($G_{TE}$) scores are almost identical to the sum of general text and emoji sentiment ($G_T + G_E$) scores from non-emoji-aware methods.

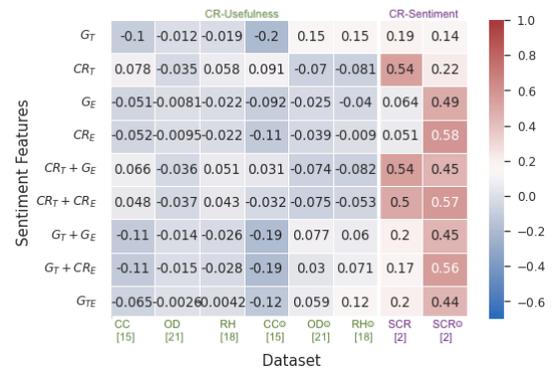

Correlation between sentiments/usefulness and features. Here, G: General, CR: Code Review, T: text sentiment, E: emoji sentiment, and 🙂: comments with emojis only

**Figure 2: Feature Values vs Usefulness and Sentiment**

Contrary to findings by Chen et al. [6], which suggested that general-domain sentiment analyzers outperform software-domain analyzers, our analysis implies that in the context of CR comments, domain-specific text & emoji sentiments ($CR_T + CR_E$) provide more meaningful information than their general-domain counterparts.

Regarding usefulness, Figure 2 reveals mixed and negligible correlations between usefulness and either text or emoji sentiments. Therefore, our first hypothesis, holds true for the 'sentiment' (**H1.1**) aspect but is invalidated for the 'usefulness' (**H1.2**) of CR comments.

Notably, these findings stem from original datasets in which only ~3% of CR comments include emojis. As a result, we investigate the impact of emojis on usefulness using an emoji-augmented dataset in our subsequent experiments concerning our second hypothesis.

## 3.2 Predicting Usefulness with Emojis (H2)

We consider both the existing feature-based and the untapped embedding-based approach to predict the CR comments' usefulness.

For the feature-based model, we examine the impact of domain-specific text and emoji sentiments (Sec 3.1), summarizing the findings in Table 2. In the context of original datasets [15, 17, 19], adding emoji features led to a modest improvement of 1.5%p in MCC scores, but only for Chromium [15] dataset. In contrast, our annotated emoji dataset, $RH^{+\text{😊}}$ and the original emoji comments from all datasets, $D_{all}^{\in \text{😊}}$, which contain emojis in CR comments, show a notable MCC increase of 7.5%p and 3.5%p respectively. This supports our second hypothesis, **H2**, emphasizing the value of emoji features in both the original and augmented datasets.

**Table 2: Performance of Predicting Usefulness with Emojis**

10-fold Stratified Cross-validation Performance Differences (Δ) between models with emojis and without emojis are reported in percentage points. Here, P: Precision, R: Recall, A: Accuracy, $F_1$: $F_1$-scores, and M: Matthews correlation coefficient (MCC)

| Dataset | Δ Feature Based | | | | | Δ Embedding Based | | | | |
| --- | --- | --- | --- | --- | --- | --- | --- | --- | --- | --- |
|  | P | R | A | M | $F_1$ | P | R | A | M | $F_1$ |
| RH [17] | 0 | 0 | 0 | 0 | 0 | 0.1 | 0.2 | 0.1 | -2.6 | 0.2 |
| CC [15] | 0.3 | 0.1 | 0.3 | **1.5** | 0.1 | -0.1 | -0.1 | -0.2 | -0.8 | -0.1 |
| OD [19] | 0 | 0 | 0.2 | 0 | -0.1 | 0.2 | 0 | 0.1 | **0.5** | -0.1 |
| $RH^{+\text{😊}}$ | 0.2 | -0.5 | 0 | 7.2 | -0.1 | 9.1 | -2.1 | 7.8 | **56.8** | 4.2 |
| $D_{all}^{\in \text{😊}}$ | 0.4 | **1.8** | **1.2** | 3.5 | **0.8** | -0.9 | -1 | -1.3 | -5.1 | -0.6 |

Turning to the embedding-based model, we observe a dramatic surge in MCC scores by 56.8%p for the augmented dataset. This method also improved precision, accuracy, and F1 score considerably. In the case of individual original datasets [15, 17, 19], the changes were not as significant, likely due to the sparse presence of emoji-containing comments. Yet, when applied to our augmented dataset, our second hypothesis, **H2**, gains strong validation through the embedding-based approach.

Interestingly, the MCC score is dropped by 5.1%p for $D_{all}^{\in \text{😊}}$, underperforming compared to three individual datasets. Since embeddings encapsulate syntactic, semantic, and other latent features, this discrepancy could indicate that the different datasets might have variations in how emojis are represented or utilized. We also see the margins of 7.2%p and 56.8%p corresponding to their emoji representation size of 1 and 300, respectively.